\documentclass[twocolumn]{jpsj2} 
\title{Thermodynamic Studies on Non Centrosymmetric Superconductors \\by AC Calorimetry under High Pressures\thanks{This paper was presented at the international workshop ``Novel Pressure-induced Phenomena in Condensed Matter Systems(NP2CMS)" August 26-29 2006, Fukuoka Japan.}\thanks{J. Phys. Soc. Jpn. {\bfseries 76}  (2007) Suppl. A, pp. 140-143}}

\author{Naoyuki \textsc{TATEIWA}$^{1}$\thanks{E-mail: tateiwa.naoyuki@jaea.go.jp}, Yoshinori \textsc{HAGA}$^{1}$, Tatsuma D. \textsc{MATSUDA}$^{1}$, Etsuji \textsc{YAMAMOTO}$^{1}$, Shugo \textsc{IKEDA}$^{1}$, \\Tetsuya \textsc{TAKEUCHI}$^{2}$, Rikio \textsc{SETTAI}$^{2}$, and Yoshichika \textsc{\=ONUKI}$^{1,2}$}

\inst{$^{1}$Advanced Science Research Center, Japan Atomic Energy Agency, Tokai, Ibaraki 319-1195, Japan \\
$^{2}$Department of Physics, Faculty of Science, Osaka University, Toyonaka 560-0043, Japan}
\recdate{July 21, 2006}

\abst{We investigated the non centrosymmetric superconductors CePt$_3$Si and UIr by the ac heat capacity measurement under pressures. We determined the pressure phase diagrams of these compounds. In CePt$_3$Si, the N\'{e}el temperature $T_{\rm N}$  = 2.2 K decreases with increasing pressure and becomes zero at the critical pressure $P_{\rm AF}$ $\simeq$ 0.6 GPa. On the other hand,  the superconducting phase exists in a wider pressure region from ambient pressure to $P_{\rm AF}$ $\simeq$ 1.5 GPa. The phase diagram of CePt$_3$Si is very unique and has never been reported before for other heavy fermion superconductors.  In UIr, the heat capacity shows an anomaly at the Curie temperature $T_{\rm C1}$ = 46 K at ambient pressure, and the heat capacity anomaly shifts to lower temperatures with increasing pressure.  The present pressure dependence of  $T_{\rm C1}$ was consistent with the previous studies by the resistivity and magnetization measurements.  Previous ac magnetic susceptibility and resistivity measurements suggested the existence of three ferromagnetic phases, FM1-3. $C_{\rm ac}$ shows a bending structure at 1.98, 2.21, and 2.40 GPa .The temperatures where these anomalies are observed are close to the phase boundary of the FM3 phase. 
}

\kword{CePt$_3$Si, UIr, superconductivity, ac calorimetry}

\begin{document}
\maketitle

\section{Introduction} 

   Recently, the discovery of non centrosymmetric superconductors such as CePt$_3$Si, UIr, CeRhSi$_3$, CeIrSi$_3$, and CeCoGe$_3$ has attracted considerable attention from both theoretical and experimental view points~\cite{bauer,akazawa1,kimura,sugitani,settai}. In these compounds, two spin degenerate bands are split due to the Rashba-type spin-orbit interaction, which strongly influences the superconducting properties, particularly the pairing symmetry of the Cooper pairs. Theoretical studies suggest a mixed-type pair function with spin triplet and singlet components~\cite{mineev}. Many theoretical and experimental studies have been extensively conducted in order to clarify this novel type of unconventional superconductivity. In this paper, we describe our thermodynamic studies on CePt$_3$Si and UIr under high pressure.

   CePt$_3$Si crystallizes in the tetragonal structure (space group $P$4$mm$) in which inversion center is absent. Superconductivity is observed at the transition temperature $T_{\rm sc}$ = 0.75 K below the N\'{e}el temperature $T_{\rm N}$  = 2.2 K at which antiferromagnetism is observed in CePt$_3$Si~\cite{bauer}.  Further, a microscopic coexistence between magnetism and superconductivity was suggested based on the neutron scattering, ${\mu}$SR and NMR experiments~\cite{metoki,amato,higemoto,yogi1}. The finding of the superconductivity led to many theoretical and experimental studies for understanding the novel superconductivity in systems without inversion centers. In the NMR experiment, two superconducting order parameters comprising the spin singlet- and triplet-pairing components have been suggested~\cite{yogi2}.  The previous high-pressure study by the resistivity measurement showed that  the superconducting transition temperature  $T_{\rm sc}$ decreases with an increase in the pressure and becomes 0 K at around 1.5 GPa~\cite{yasuda1}.  On the other hand, the pressure dependence of the antiferromagnetic transition temperature $T_{\rm N}$ is not clear because the anomaly in the resistivity at $T_{\rm N}$ is too weak to be detected. 

   UIr is a ferromagnetic compound with the Curie temperature $T_{\rm C1}$ = 46 K~\cite{dommann,galatanu}. UIr crystallizes in the monoclinic PbBi-type structure (space group $P$2$_{1}$). There is no inversion center in the crystal structure.  The magnetic susceptibility obeys the Curie-Weiss law at temperatures above 500 K, with an effective magnetic moment $\mu_{\rm eff}$ = 3.6 $\mu_{\rm B}$/U; this value corresponds to the free-ion value of the localized 5$f^2$ and/or 5$f^3$ configurations. In the ferromagnetic state, the magnetization is highly anisotropic and the magnetic property is of the Ising type.  The size of the spontaneous magnetic moment oriented along the [10\=1] direction is 0.5 $\mu_{\rm B}$/U. The cyclotron effective mass in the range of 10 - 30$m_{0}$ was observed in the de Haas van Alphen experiment. The electronic specific heat coefficient was determined to be $\gamma$ = 49 mJ/K$^2$${\cdot}$mol \cite{yamamoto1,yamamoto2}. The itinerant character of the 5{\it f} electrons at low temperatures was suggested.  Previous high-pressure studies showed that  the Curie temperature $T_{\rm C1}$ decreases  with increasing pressure and that superconductivity is observed in a narrow pressure region from 2.6 to 2.7 GPa\cite{edbauer,akazawa1,akazawa2}. A recent study revealed the existence of three ferromagnetic phases FM1-3 and that the superconducting phase exists at a pressure just below the critical pressure of the FM3 phase~\cite{kobayashi}.  
    
  In order to establish the pressure phase diagram of both the compounds,the thermodynamic measurements are necessary. We carried out the ac heat capacity measurement on CePt$_3$Si and UIr under high pressures. We present the experimental results in this paper.
  
\section{Experimental}   
   A single crystal of CePt$_3$Si was grown by the Bridgeman method and that of UIr was grown by the Czochralski method in a tetra-arc furnace. The details of the sample preparation are given in our previous papers \cite{yasuda1,yamamoto1}. These values of the residual resistivity ratio RRR (= ${\rho}_{\rm RT}/{\rho}_{\rm 0}$) are 100 and 200 for CePt$_3$Si and UIr, respectively. Those values indicate the high quality of the single crystal samples. The ac heat capacity measurement under pressure was measured using a AuFe-Au thermocouple in a hybrid piston cylinder-type cell. The details of experimental techniques for the ac calorimetry are given in ref. [20].    

\section{Results and discussions} 
  \subsection{CePt$_3$Si}
\begin{figure}[b]
\begin{center}
\includegraphics[width=8.1cm]{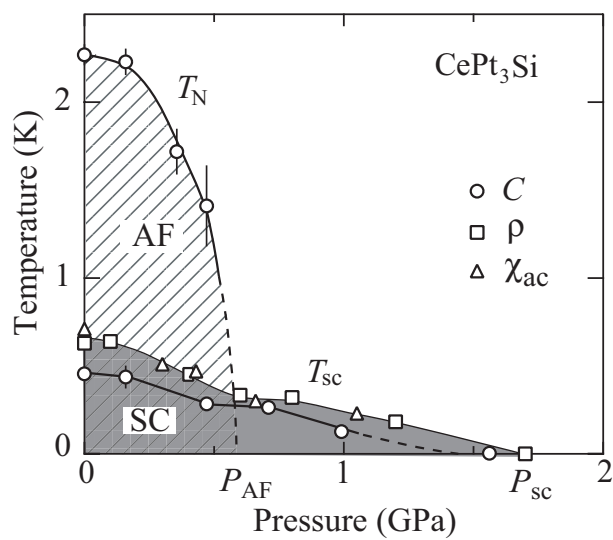}
\end{center}
\caption{Pressure phase diagram of CePt$_3$Si.}
\label{f1}
\end{figure}

  The pressure-temperature phase diagram of CePt$_3$Si obtained by the heat capacity measurement is shown in Figure 1.  The superconducting  transition temperatures  $T_{\rm sc}$ previously determined by both resistivity $\rho$ and ac susceptibility $\chi_{\rm ac}$ measurements are also plotted in the figure~\cite{yasuda1}. The N\'{e}el  temperature determined by the resistivity measurement was not plotted because there was ambiguity in the determination of $T_{\rm N}$ due to the weak resistivity anomaly at the transition temperature. With increasing pressure, $T_{\rm N}$ decreases faster than $T_{\rm sc}$ and becomes 0 K at around $P_{\rm AF}$ $\simeq$ 0.6 GPa. $T_{\rm sc}$ decreases with increasing pressure and becomes approximately constant from 0.6 GPa to 0.8 GPa. It decreases furthermore with increasing pressure and becomes zero at $P_{\rm sc}$ $\simeq$ 1.5 GPa. Thus, the pressure dependence of $T_{\rm sc}$ shows a characteristic feature. The present pressure of 0.6 GPa corresponds to the antiferromagnetic critical pressure $P_{\rm AF}$.  CePt$_3$Si is in the paramagnetic state from $P_{\rm AF}$ $\simeq$ 0.6 GPa to 1.5 GPa and it shows only superconducting transitions.  
  
  Pressure-induced superconductivity was observed in several cerium compounds such as CeIn$_3$, CeRh$_2$Si$_2$ and CePd$_2$Si$_2$~\cite{flouquet}. These compounds show the antiferromagnetic ordering at ambient pressure.  The $T_{\rm N}$ value decreases with increasing pressure. Superconductivity is observed around the magnetic critical point $P_{\rm AF}$, where $T_{\rm N}$ becomes zero. The superconducting phase exists in a narrow pressure region around the antiferromanetic critical pressure $P_{\rm AF}$. The $T_{\rm sc}$ value becomes maximum around the critical pressure. Superconductivity is considered to be mediated by the low-energy magnetic excitations around the magnetic quantum critical point$P_{\rm AF}$. On the other hand, in the case of CePt$_3$Si,  the bulk superconducting phase exists in a wide pressure region above and below $P_{\rm AF}$, and $T_{\rm sc}$ does not shows a maximum at $P_{\rm AF}$. The maximum $T_{\rm sc}$ value is realized at ambient pressure.  From the pressure dependence of the linear heat capacity coefficient $\gamma$, it was suggested that the critical pressure $P_{\rm AF}$ is not of the second-order quantum critical point but that of the first-order~\cite{tateiwa}. Therefore, the superconductivity in CePt$_3$Si may be different from that around the magnetic quantum critical point. 

  \begin{figure}[b]
\includegraphics[width=8.8cm]{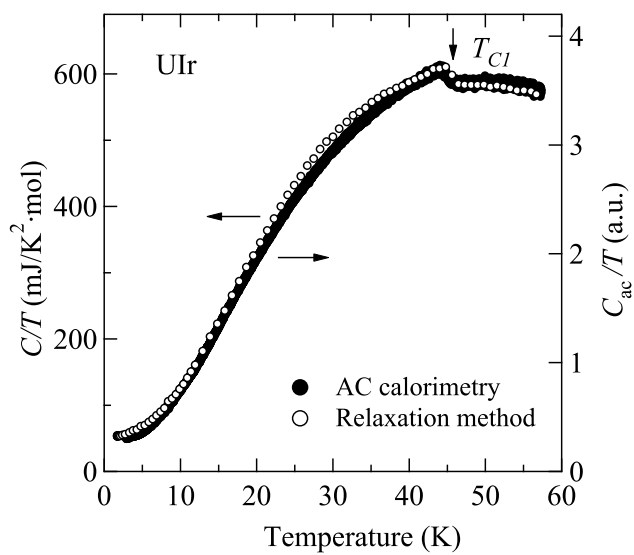}
\caption{Temperature dependence of $C_{\rm ac}$ by the present measurement (right side) and the data obtained by the relaxation method (left side). }
  	\label{fig-1}
  \end{figure}

\begin{figure}[t]
\begin{center}
\includegraphics[width=8.7cm]{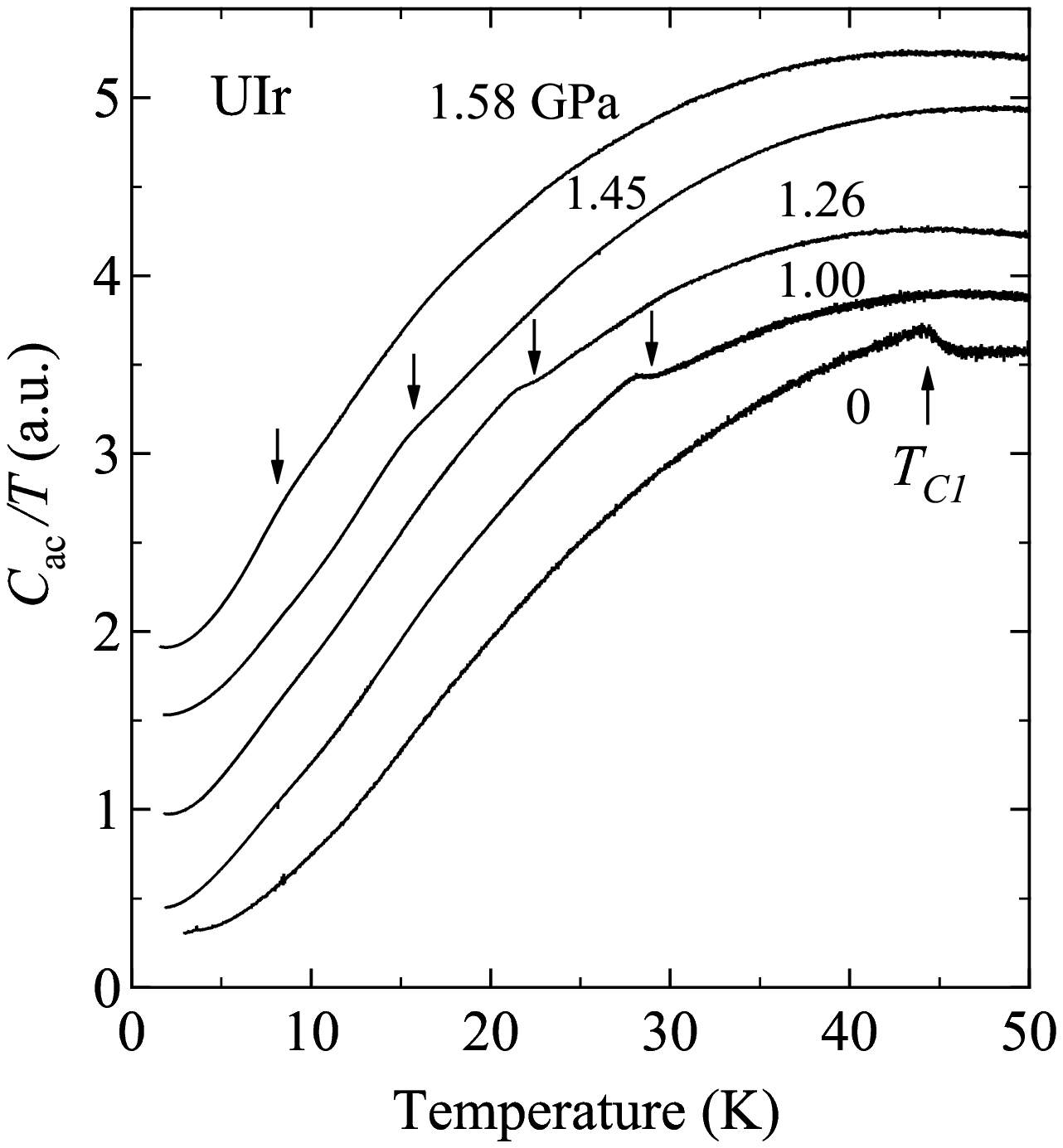}
\end{center}
\caption{Temperature dependence of the heat capacity of UIr under high pressures.}
\label{f1}
\end{figure}

\begin{figure}
\begin{center}
\includegraphics[width=8.7cm]{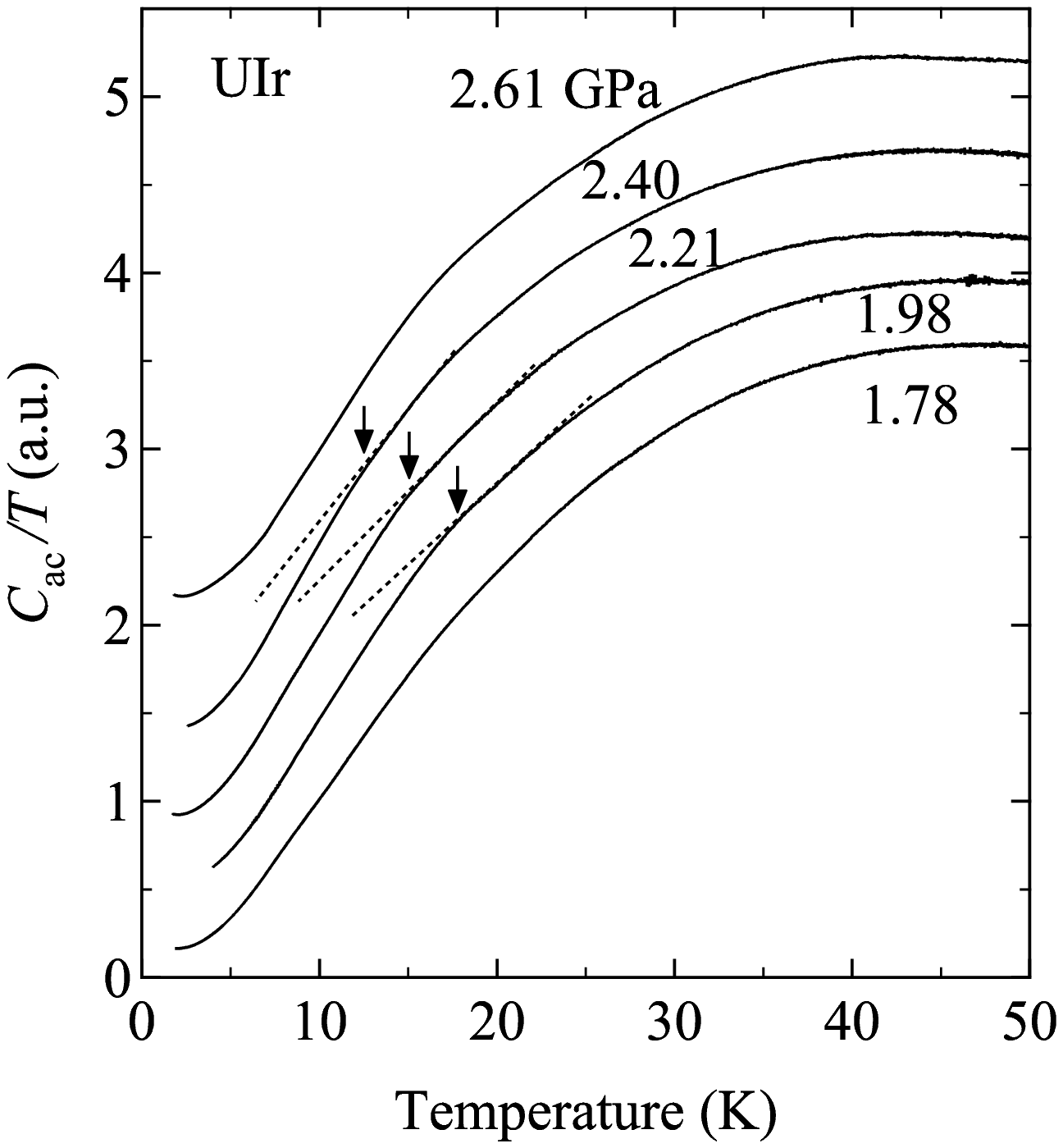}
\end{center}
\caption{Temperature dependence of the heat capacity of UIr under high pressures.}
\label{f1}
\end{figure}

\begin{figure}[t]
\begin{center}
\includegraphics[width=8cm]{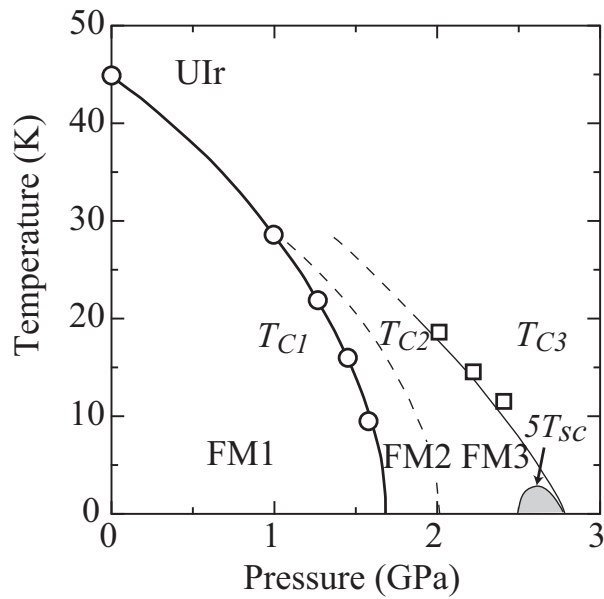}
\end{center}
\caption{Pressure phase diagram of UIr}
\label{f1}
\end{figure}

  \subsection{UIr}
   Figure 2 shows the temperature dependence of $C_{\rm ac}$ (right side) at ambient pressure. The experimental data obtained by the relaxation method is also plotted(left side). Both the data are scaled around 50 K by experimental data. The temperature dependence of $C_{\rm ac}$ is qualitatively consistent with that obtained by the relaxation method. At ambient pressure, the heat capacity shows an anomaly at around the ferromagnetic transition temperature $T_{\rm C1}$ = 46 K. 

  Figure 3 shows the heat capacity $C_{\rm ac}$ under pressures up to 1.58 GPa. The experimental data under pressures are simply shifted upwards.  With increasing pressure, the anomaly at $T_{\rm C1}$ shifts to the lower temperature side and the strength of the anomaly becomes weaker. At 1.58 GPa, a weak anomaly is observed at around 10 K. 
         
  \begin{table}[b]
\caption{Calculated heat capacity jump}
\begin{center}
\setlength{\tabcolsep}{6pt}
\begin{tabular}{ c | c | c | c   } \hline
$P$(GPa) & $T_{m}$ (K) & ${\Delta}C/T_{m}^{\rm (SCR)}$ & ${\Delta}C/T_{m}^{\rm (Stoner)}$  \\ 
  &  & (mJ/K$^2$${\cdot}$mol) & (mJ/K$^2$${\cdot}$mol)\\ \hline
0 & $T_{C1}$ = 46 K & 11 & 44\\ 
1.8  & $T_{C2}$ = 13 K & 0.11 & 0.44\\ 
2.2 & $T_{C3}$ = 14 K & 0.22 & 0.88\\ 
2.4 & $T_{C3}$ = 10 K & 0.30 & 1.2\\ 
2.6 & $T_{C3}$ = 5 K & 0.6 & 2.4\\ \hline
\end{tabular}
\end{center}
\label{tab:moments}
\end{table}%
  Figure 4 shows the heat capacity $C_{\rm ac}$ in the pressure region from 1.58 GPa to 2.61 GPa. The experimental data under pressures are simply shifted upwards. In this region, there is no distinct heat capacity anomaly. However the curve of $C_{\rm ac}$ at 1.98, 2.21 and 2.40 GPa shows a bending structure as shown in Figure. 4. In the Fig.4, arrows are used to indicate the temperature at which $C_{\rm ac}$ shows bending. As discussed later, the temperatures are close to the phase boundary of the FM3 phase. No anomaly is observed in $C_{\rm ac}$ around the phase boundary of the FM2.

   The pressure phase diagram of UIr is shown in Figure 5.  The transition temperatures $T_{\rm C1}$ obtained from the present study are indicated by circles. The phase boundaries indicated by solid lines are determined by the resistivity and dc magnetization measurements; the ones indicated by broken lines are determined by ac susceptibility measurements~\cite{kobayashi}. The superconducting region is indicated by a shadow at an enlarged scale.  It is found that the pressure dependence of $T_{\rm C1}$ determined by the present heat capacity measurement is consistent with those determined in the previous study. The temperatures where $C_{\rm ac}$ shows bending are indicated by squares. Interestingly, these temperatures are close to the boundary of the FM3 phase. Thus, the bending anomaly may be the thermodynamic one related of the phase boundary of the FM3 phase. The anomaly does not appear in $C_{\rm ac}$ at 2.61 GPa where the value of $T_{\rm C3}$ is estimated to be approximately 5 K from the previous studies~\cite{kobayashi}. The reason for the absence of the heat capacity anomaly is not clear. One possibility is that the critical pressure of the FM3 phase in the present study may be different from that of previous reports due to sample dependence or ambiguity in pressure  determination.

  Next, we discuss the magnitude of the heat capacity anomaly in UIr. The heat capacity anomaly in itinerant ferromagnets has been the subject of theoretical studies\cite{wohlfarth, moriya,murata,lonzarich}. In the Stoner-Wohlfarth model, the jump of the heat capacity at the transition temperature was estimated as ${\Delta}C/T_{m}^{\rm (Stoner)}$ = ($\mu_0M^2_0$)/($\chi_0T_{m}^2$), where $M_0$ is the spontaneous magnetization and $\chi_0$ is the magnetic susceptibility in the ordered state. The Stoner theory does not include the effect of the spin fluctuations and the heat capacity jump is overestimated. Mohn and Hilscher derived the following expression for the heat capacity jump at the transition temperature within the frame work of the self-consistent renormalized (SCR) spin fluctuation theory based on the model by Murata and Doniach, ${\Delta}C/T_{m}^{\rm (SCR)}$ = ($\mu_0M^2_0$)/(4$\chi_0T_{m}^2$)\cite{mohn}.  It should be noted that the value of ${\Delta}C/T_{m}^{\rm (SCR)}$ is 1/4 that of ${\Delta}C/T_{m}^{\rm (Stoner)}$. Mohn and Hilscher compared the experimental results of several itinerant ferromagnets with the theoretical models and found that the experimental values lay between ${\Delta}C/T_{m}^{\rm (SCR)}$ and ${\Delta}C/T_{m}^{\rm (Stoner)}$, an upper and lower bounds\cite{mohn}.
   
   In the case of UIr, the value of ${\Delta}C/T_{m}^{\rm (SCR)}$ at ambient pressure is calculated to be 11 mJ/K$^2$${\cdot}$mol, using $M_0$ = 1.1 $\times$ 10$^5$ A/m and $\chi_0$ = 4.0 $\times$ 10$^{-5}$, which are determined from the magnetization measurement at 1.8 K~\cite{galatanu}. From the present experiment, ${\Delta}C/T_{m}^{\rm (exp.)}$ is estimated to be 25 mJ/K$^2$${\cdot}$mol. This value lies between ${\Delta}C/T_{m}^{\rm (SCR)}$ and ${\Delta}C/T_{m}^{\rm (Stoner)}$. The calculated values under pressures are listed in Table 1. In the calculations, we used the values of $M_0$ and $\chi_0$ from the previous magnetization measurements under pressure~\cite{akazawa2}. The spontaneous magnetic moments of the FM2 and FM3 phases are very small, approximately 0.05 $\mu_{\rm B}$/U. The calculated value of ${\Delta}C/T_{m}$ is significantly reduced in the FM2 and FM3 phases. This may be the reason that $C_{\rm ac}$ does not show a clear heat capacity jump but shows  bending structure at  $T_{\rm C3}$.

\section{Summary}
 We present our experimental results on the nonconetrosymetric superconductors CePt$_3$Si  and UIr. The pressure phase diagrams of both these compounds are determined from the thermodynamic viewpoint. In CePt$_3$Si, the antiferromagnetic phase disappears at $P_{\rm AF}$ $\simeq$ 0.6 GPa. Meanwhile, the superconducting phase exists in a wider pressure region from the ambient pressure to $P_{\rm AF}$ $\simeq$ 1.5 GPa. The phase diagram of CePt$_3$Si is very unique. The superconductivity appears to be different  from those observed around the magnetic quantum critical point. In UIr, the heat capacity anomaly was observed at $T_{\rm C1}$ and the pressure dependence of  $T_{\rm C1}$ by the present study was consistent with that obtained in previous studies by the resistivity and magnetization measurements.  $C_{\rm ac}$ shows bending at around $T_{\rm C3}$. The weakness in the heat capacity anomaly may be due to the very small ordered moment of the FM3 phase.

\section{Acknowledgements}
 This work was financially supported by the Grant-in-Aid for Creative Scientific Research (15GS0213), Scientific Research of Priority Areas and Scientific Research (A) from the Ministry of Education, Culture, Sports, Science and Technology (MEXT).

\end{document}